\documentclass[twocolumn,aps,pre,superscriptaddress,floatfix]{revtex4-2}
\usepackage{graphicx, graphics, amsmath, amssymb, rotate, textcomp, gensymb}
\usepackage{mathrsfs, float}
\usepackage[T1]{fontenc}
\usepackage{dcolumn}
\usepackage{physics}
\usepackage{wrapfig}
\usepackage[colorlinks]{hyperref}
\usepackage{xcolor}
\usepackage[none]{hyphenat}
\usepackage{times}
\usepackage{natbib}
\usepackage{multirow}

\begin{document}

\title{Finite-size Scaling in Kinetics of Phase Separation in Certain Models of Aligning Active Particles}
\author{Tanay Paul}
\affiliation{Theoretical Sciences Unit and School of Advanced Materials, Jawaharlal Nehru Centre for Advanced Scientific Research, Jakkur, Bangalore 560064, India}
\author{Nalina Vadakkayil}
\affiliation{Theoretical Sciences Unit and School of Advanced Materials, Jawaharlal Nehru Centre for Advanced Scientific Research, Jakkur, Bangalore 560064, India}
\affiliation{Complex Systems and Statistical Mechanics, Department of Physics and Materials Science, University of Luxembourg, Luxembourg, L-1511, Luxembourg}
\author{Subir K. Das}
\email{das@jncasr.ac.in}
\affiliation{Theoretical Sciences Unit and School of Advanced Materials, Jawaharlal Nehru Centre for Advanced Scientific Research, Jakkur, Bangalore 560064, India}
\date{\today}

\begin{abstract}
To study the kinetics of phase separation in active matter systems, we consider models that impose a Vicsek-type self-propulsion rule on otherwise passive particles interacting via the Lennard-Jones potential.
Two types of kinetics are of interest: one conserves the total momentum 
of all the constituents and the other that does not.
We carry out numerical simulations, assisted by molecular dynamics, 
to obtain results on structural growth and aging properties. 
Results from our studies, with various finite boxes, 
show that there exist scalings with respect to the system sizes, 
in both quantities, as in the standard passive cases. 
We have exploited this scaling picture to accurately estimate 
the corresponding exponents, in the thermodynamically large system size limit, 
for power-law time-dependences. 
It is shown that certain analytical functions 
describe the behavior of these quantities quite accurately, including the finite-size limits. 
Our results demonstrate that even though the conservation of velocity has at best weak effects 
on the dynamics of evolution in the thermodynamic limit, 
the finite-size behavior is strongly influenced 
by the presence (or the absence) of it.  

\end{abstract}
\maketitle

\section{Introduction}\label{sec:intro}

\sloppy Active matter systems contain objects that can self-propel by drawing 
energy from the surroundings \cite{marchettirev, ramaswamyrev, binderrev, daschapter, gompperrev, bechingerrev}. 
Examples of such inherently out-of-equilibrium systems cover a rather wide range of 
length-scales \cite{ramaswamyrev, binderrev, daschapter}, in terms of the size of the 
constituents as well as the space that they collectively occupy. 
Interesting structures and dynamics that are observed in these systems also span a 
large range of length as well as time. 
These can be seen in small systems like a cytoskeletal network \cite{ramaswamyrev, surrey01} 
or a bacterial colony \cite{dombrowski04} as well as in systems as big as a school of fish \cite{lopez12}, 
a flock of birds \cite{cavagna14} and a human assembly \cite{pince16}. 
Despite the fact that the constituents of the systems are vastly different, one may 
expect existence of universality of sorts similar to passive systems 
\cite{binderrev, daschapter}. 
The nonequilibrium phase transitions that these systems undergo often
involve a switch from a disordered state to a state having a coherent motion of the 
constituents, with long-range directional order, and consequent clustering. 
These frequently bear characteristics similar to those observed in equilibrium phase 
transitions associated with passive matter \cite{binderrev, daschapter}. 
This is despite the fact that instead of attaining a thermal equilibrium these 
effectively driven systems approach a steady state in late time. 
Understanding the kinetics concerning such approaches, 
which include structural growth and aging, 
as a transition occurs from one state to the other, 
is of practical as well as fundamental importance 
\cite{binderrev, daschapter, das17, belmonte08, peruani13, cremer14}. 
Before stating the specific objective and importance of the work, 
below we briefly describe the key concepts of evolutionary dynamics 
that are established via the studies of passive matter systems.

At a given point in time, information about many aspects of the structure of a system 
can be obtained by calculating the two-point equal-time ($t$) correlation function \cite{bray02},
\begin{equation}
    C(r,t)= \langle (\psi(\vec{r}_1,t)-\langle \psi(\vec{r}_1,t) \rangle) (\psi(\vec{r}_2,t)-\langle \psi(\vec{r}_2,t) \rangle) \rangle ,\label{eq:cr}
\end{equation}
where $r=|\vec{r}_1-\vec{r}_2|$, $\vec{r}_1$ and $\vec{r}_2$ being two positions in space, 
and $\psi$ is a time dependent local order parameter that can, e.g., 
be a concentration or a velocity direction, depending upon the type of the transition. 
Here $\langle ... \rangle$ represents statistical averaging. 
When the local concentration of particles is chosen to be equal to $\psi$, 
$C(r,t)$ in Eq. \eqref{eq:cr} represents the correlation in density variation 
as a function of $r$ at a given time ($t$). 
Often in an evolving system, following a quench, such correlation increases with time, 
implying growth of particle-rich and particle-poor domains, 
typically in a self-similar way \cite{bray02}. 
The latter means, with passing time, the average size of domains, $\ell(t)$, 
that can be quantified from the decay of $C(r,t)$, grows 
but the structure at any instant remains statistically similar to that at an earlier time. 
Mathematically, this scaling feature is expressed as \cite{bray02}
\begin{equation}
    C(r,t)\equiv \hat{C}\left(\frac{r}{\ell(t)}\right). \label{eq:crScaling}
\end{equation}
Here $\hat{C}$ is a function independent of time. 
The time-dependence of $\ell$ is typically of power-law type \cite{bray02}, viz.,
\begin{equation}
    \ell(t) \sim t^\alpha, 
    \label{eq:growth}
\end{equation}
$\alpha$ being referred to as a growth exponent. 
On the other hand, unlike the domain growth, 
which can be quantified via a single-time correlation function, 
the aging behavior of the system \cite{dfisher88}, another crucial dynamic aspect, 
is typically studied via the 
the two-time autocorrelation function \cite{dfisher88}:
\begin{equation}
    C_{\rm{ag}}(t,t_w)= \langle (\psi(\vec{r},t)-\langle \psi(\vec{r},t) \rangle) (\psi(\vec{r},t_w)-\langle \psi(\vec{r},t_w) \rangle) \rangle, \label{eq:cag}
\end{equation}
where $t_w(\leq t)$, the age of the system, is referred to as the waiting time. 
This function captures the correlation between two configurations, 
separated in time by $(t-t_w)$. 
A slower decay of $C_{\rm{ag}}(t,t_w)$ with increasing $t_w$ 
is characteristic of the aging phenomena. 
While $C_{\rm{ag}}(t,t_w)$ does not obey the time translation invariance in a typical evolving system, 
its decay may exhibit the dynamical scaling \cite{dfisher88, das17},
\begin{equation}
    C_{\rm{ag}}\equiv \hat{C}_{\rm{ag}}\left( \frac{\ell}{\ell_w} \right), \label{eq:cagScaling}
\end{equation}
$\ell_w$ being the characteristic length-scale at $t=t_w$. 
Often it is found that $C_{\rm{ag}}$ obeys a power-law behavior 
in the long time limit \cite{dfisher88}:
\begin{equation}
    C_{\rm{ag}}\sim \left( \ell/\ell_w \right)^{-\lambda}, \label{eq:agingexp}
\end{equation}
$\lambda$ being the aging exponent. 
The values of the exponents $\alpha$ and $\lambda$ 
usually have dependence on the transport as well as morphological features
\cite{roy13JCP, binder76}. 

In several passive matter systems such scaling properties have been identified and 
the exponents have been estimated 
to obtain information on universality in evolution dynamics. 
Efforts have started in the case of active matter systems \cite{das17, janssen17}. 
In the theoretical literature, simple models have been constructed 
to reproduce interesting experimental observations in this subdomain. 
One such model is referred to as the Vicsek model (VM) \cite{vicsek95, vicsek97} 
that generates interesting collective motion of the constituents, mimicking real physical systems. 
Within the framework of this model, a point-like individual tries 
to orient its motional direction along the average velocity of its neighbors 
lying within a certain range. 
At low strengths of imposed noise and high number densities, 
these individual particles form clusters.
Transition to such a state, starting from a homogeneous configuration, resembles 
typical vapor-liquid transition in a passive situation 
\cite{baglietto09, gregoire04, solon2015phase}. 
In recent times, variants of this model are introduced 
by incorporating more realistic features like the size and shape of an object 
\cite{das17, gregoire01, gregoire03, gregoire04, chate08}. 
These require the addition of passive interactions, 
to prevent inter-particle overlap. 
For such a composite model, if the passive limit exhibits a phase transition, 
one derives an advantage in terms of 
quantifying the effects of the activity by drawing a comparison with the former. 
It is then interesting to ask, how with the variation of active and passive 
interactions the scaling properties with respect to structure, 
growth, and aging change as a system evolves to a steady state.

It is important to note here that the above-discussed scaling pictures are 
strictly true in the thermodynamically large system size limit. 
Deviations are expected when the sizes are finite. 
Studies in such finite boxes should be of particular interest 
in the case of active matter, given that the Avogadro number of constituents, 
in this case, does not exist. 
Keeping this in mind, in this work, we study growth and aging 
in model active matter systems of finite sizes. 
For this purpose, we combine the VM \cite{vicsek95} 
with a passive Lennard-Jones model \cite{allenbook}, 
both independently providing phase transitions in the appropriate limiting situations. 
We demonstrate that finite-size scaling exists in both the quantities. 
We exploit \cite{chate08} this scaling to obtain results in the limits 
if the sizes were thermodynamically large. 
The latter is also of much fundamental academic importance.
Furthermore, we verify the effects of the global conservation of velocity, 
a key field in active matter dynamics.  
Note that there exists such a model variant in the literature of passive matter also,
with the global conservation of order parameter \cite{annett92, sire95}.
Another crucial aspect of our study is related to the description of the growth and aging 
via certain analytical functions combining both thermodynamic and finite-size limit features.

The organization of the rest of the manuscript is as follows. 
In section \ref{sec:model}, the models and methods used in this work are described. 
Section \ref{sec:results} contains the results and analyses. 
Finally, section \ref{sec:conclusion} summarizes the paper with a future outlook.

\section{Model and Methods}\label{sec:model}

Our systems consist of spherical particles of mass $m=1$. 
A particle interacts with all other particles within a spherical region, 
having radius $r_c$, around it, via the passive pair potential \cite{roy13JCP, midya17},
\begin{equation}
    V(r)= \tilde{V}(r)-\tilde{V}(r_c)-(r-r_c)\left( \dv{\tilde{V}}{r}\right)_{r=r_c}. \label{eq:passive}
\end{equation}
Here, $r$ is the separation between two particles 
and $\tilde{V}(r)$ is the Lennard-Jones (LJ) potential \cite{allenbook},
\begin{equation}
    \tilde{V}(r)= 4\epsilon\left[ \left( \frac{\sigma}{r} \right)^{12} - \left( \frac{\sigma}{r} \right)^{6} \right]. \label{eq:LJ}
\end{equation}
In Eq. \eqref{eq:LJ}, $\sigma$ is the inter-particle diameter 
and $\epsilon$ is the interaction strength between a pair. 
We have chosen $r_c=2.5\sigma$. 
The phase behavior of this passive system is well studied in different dimensions $d$. 
For a vapor-liquid transition, the critical values of temperature and density 
\cite{errington03, midya17} in $d=3$ are $T_c\simeq 0.94\epsilon/k_B$ and $\rho_c\simeq 0.32$, 
$k_B$ being the Boltzmann constant.

The self-propulsion rule, in our study, is very similar to the VM. 
Here also, like in the VM, the velocity of the $i$-th particle is influenced by 
the average direction, $\vec{D}_i$, of all the particles, including the $i$-th particle, 
within a spherical cut-off region. 
In our model, this active interaction range is the same as $r_c$. 
The average direction $\vec{D}_i$ is calculated as \cite{das17}
\begin{equation}
    \vec{D}_i=\frac{\sum\limits_{j\in \mathscr{R}_c}\vec{v}_j}{\bigg|\sum\limits_{j\in \mathscr{R}_c}\vec{v}_j\bigg|}, \label{eq:Di}
\end{equation}
where $\vec{v}_j$ is the velocity of the $j$-th particle 
and the summation is carried over all particles inside 
the sphere $\mathscr{R}_c$ of radius $r_c$ around the $i$-th particle. 
This direction $\vec{D}_i$ influences the velocity of the $i$-th particle 
in such a way that its velocity magnitude remains unchanged, like in the VM, 
only its direction of motion changes. 
The corresponding update rule will be discussed later.

Our MD simulations, in Canonical ensemble, have been carried out 
in periodic cubic simulation boxes of linear sizes $L\sigma$. 
The temperature of the system has been controlled by using a Langevin Thermostat. 
Thus, in absence of an active interaction, 
the equation of motion for the $i$-th particle has the form \cite{frenkelbook}
\begin{equation}
    m\ddot{\vec{r}}_i=-\vec{\nabla}V_i-\gamma m \dot{\vec{r}}_i+\sqrt{6m\gamma k_B T}\vec{R}_i(t). \label{eq:eom}
\end{equation}
In Eq. \eqref{eq:eom}, the term $-\vec{\nabla}V_i$ is the force on the $i$-th particle 
due to the passive interaction given in Eq. \eqref{eq:passive}, 
$\gamma$ is a damping coefficient and $\vec{R}_i(t)$ is a random noise, 
having components drawn from a uniform distribution within the range $[-1,1]$ such that
\begin{equation}
    \langle R_{i\nu}(t)R_{j\mu}(t^\prime) \rangle=\delta_{ij}\delta_{\nu\mu}\delta(t-t^\prime). \label{eq:deltacorr}
\end{equation}
Here $i$, $j$ are the particle indices; 
$\nu$, $\mu$ are indices that indicate Cartesian directions; 
and $t$, $t^\prime$ are different times. 
The coefficient of the noise vector $\vec{R}_i$ is connected to the drag term 
via the Fluctuation-Dissipation relation \cite{kubo66}. 
Eq. \eqref{eq:eom} is solved using the Velocity Verlet algorithm 
\cite{allenbook, frenkelbook} to obtain the velocity 
$\vec{v}_i^{\textrm{\,pas}}(t+\Delta t)$, 
$\Delta t$ being the integration time-step, at time $t+\Delta t$, from $t$.
Following this, $\vec{v}_i^{\textrm{\,pas}}(t+\Delta t)$ is updated further 
using $\vec{D}_i$ that incorporates the effects of self-propulsion. 
This is described below.

As prescribed in Ref.\cite{das17}, an active force can be defined as
\begin{equation}
    \vec{f}_i=f_a\vec{D}_i, \label{eq:activeForce}
\end{equation}
where the strength $f_a$ is a constant. 
This has been used to obtain the velocity of the $i$-th particle from 
$\vec{v}_i^{\textrm{\,pas}}(t+\Delta t)$ to
\begin{equation}
    \vec{v}_i^{\,\prime}(t+\Delta t)=\vec{v}_i^{\textrm{\,pas}}(t+\Delta t)+\frac{\vec{f}_i}{m}\Delta t. \label{eq:addActivity}
\end{equation}
The force $\vec{f}_i$ not only changes the direction but also the magnitude of velocity, 
which may lead to a change in the kinetic temperature, despite the thermostat. 
To keep the latter constant, at an assigned value, 
the final velocity of the $i$-th particle at $(t+\Delta t)$ has been calculated as
\begin{equation}
    \vec{v}_i(t+\Delta t)=\frac{|\vec{v}_i^{\textrm{\,pas}}(t+\Delta t)|}{|\vec{v}_i^{\,\prime}(t+\Delta t)|}\vec{v}_i^{\,\prime}(t+\Delta t). \label{eq:velUp}
\end{equation}
For the case of the conserved dynamics, 
after this step, to make the total momentum zero, 
the average velocity of all the particles 
has been subtracted from the velocity of each particle. 
In the simulations of Ref. \cite{das17}, $\vec{f}_i$ contained 
the influence of the magnitude of average velocity of the neighbors as well. 
There, for certain computational convenience the unit of time was chosen to be 
$t_0^\prime = \sqrt{m\sigma^2/48\epsilon}$, 
whereas in this work the unit is $t_0=\sqrt{m\sigma^2/\epsilon}$. 
Furthermore, we have $\sigma$, $\epsilon$, and $\epsilon/k_B$ as the units of length, energy and temperature, respectively. 
For the integration time step, we have used $\Delta t=0.005t_0$. All the results presented in this paper are produced with $\gamma=1$ and $f_a=1$.

Configurations with particles placed randomly inside the simulation boxes, 
with number density $\rho_0=0.3$, have been considered as initial configurations 
which are quenched to the temperature $T=0.5$. 
We have studied systems with sizes $L=50$, $64$, $80$, and $100$. 
A Finite-size scaling (FSS) approach has been incorporated for calculating $\alpha$ and $\lambda$. 
Note that for density $\rho_0=0.3$, a system of size $L=100$ contains $N=300000$ particles. 
For all the system sizes, we have considered averaging over a minimum of $40$ 
independent initial configurations for obtaining the final quantitative results. 
All the results presented here are for $L=100$, 
except when specific FSS analyses are carried out.
As opposed to Ref. \cite{das17}, 
here the objective is to demonstrate and quantify the finite-size scaling behavior, 
alongside obtaining accurate information on the evolution 
for both conserved and nonconserved velocity dynamics. 
It is important to note here that quantification of finite-size behavior 
in active matter systems for the full range 
was not done previously, to the best of our knowledge. 
Furthermore, for such systems this is the first FSS study for the aging dynamics. 

\section{Results}\label{sec:results}
\begin{figure}[b]
    \centering
    \includegraphics[width=0.48\textwidth]{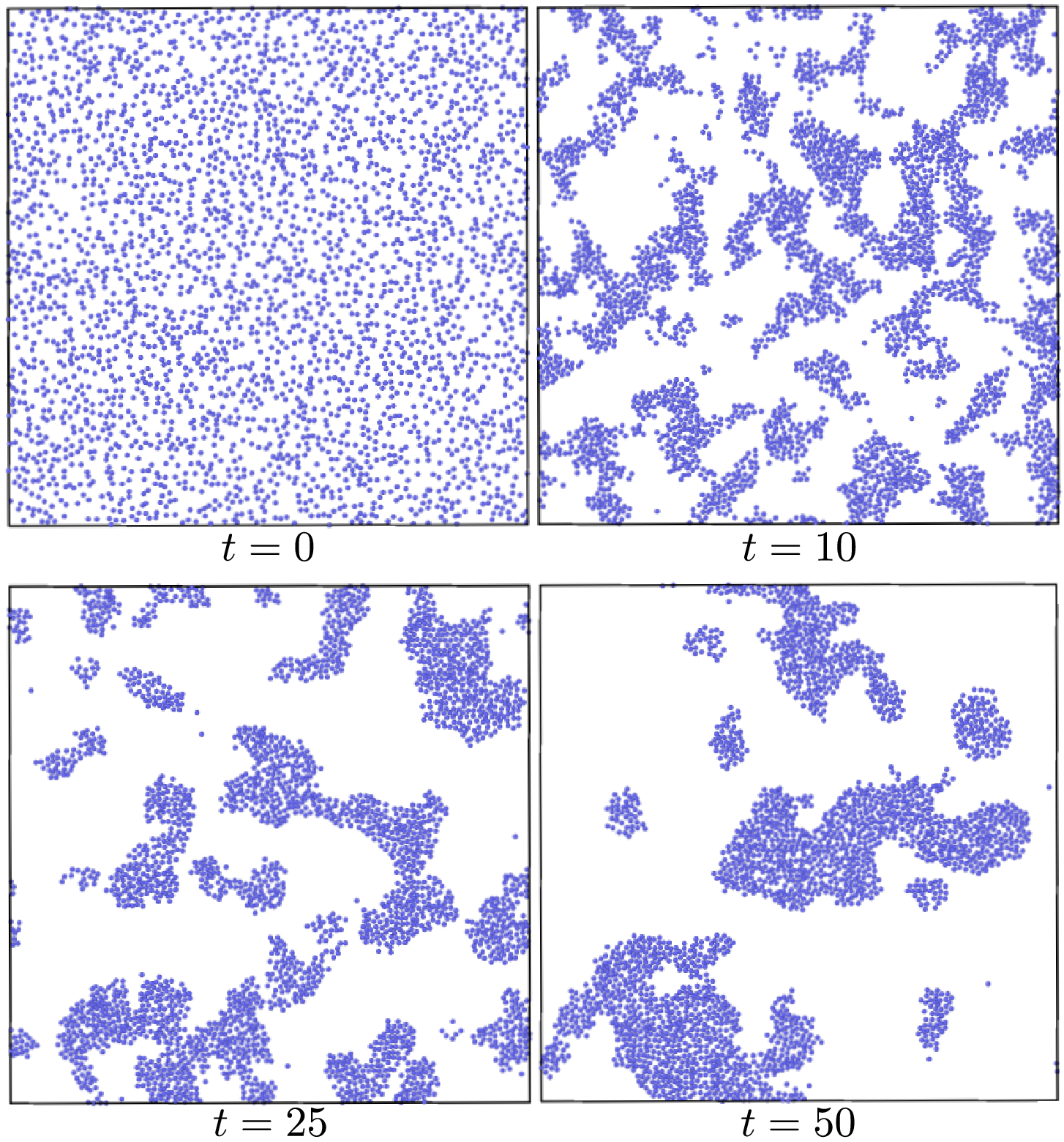}
    \caption{Two-dimensional cross-sections of the evolution snapshots from a computer experiment with the considered model with conserved velocity. The pictures were recorded following a quench of a homogeneous configuration to $T=0.5$, after setting $f_a$ at unity. Here the linear dimension of the box is $L=100$, containing $N=300000$ particles. The locations of the particles are marked.}
    \label{fig:layerview}
\end{figure}
\begin{figure}[t]
    \centering
    \includegraphics[width=0.48\textwidth]{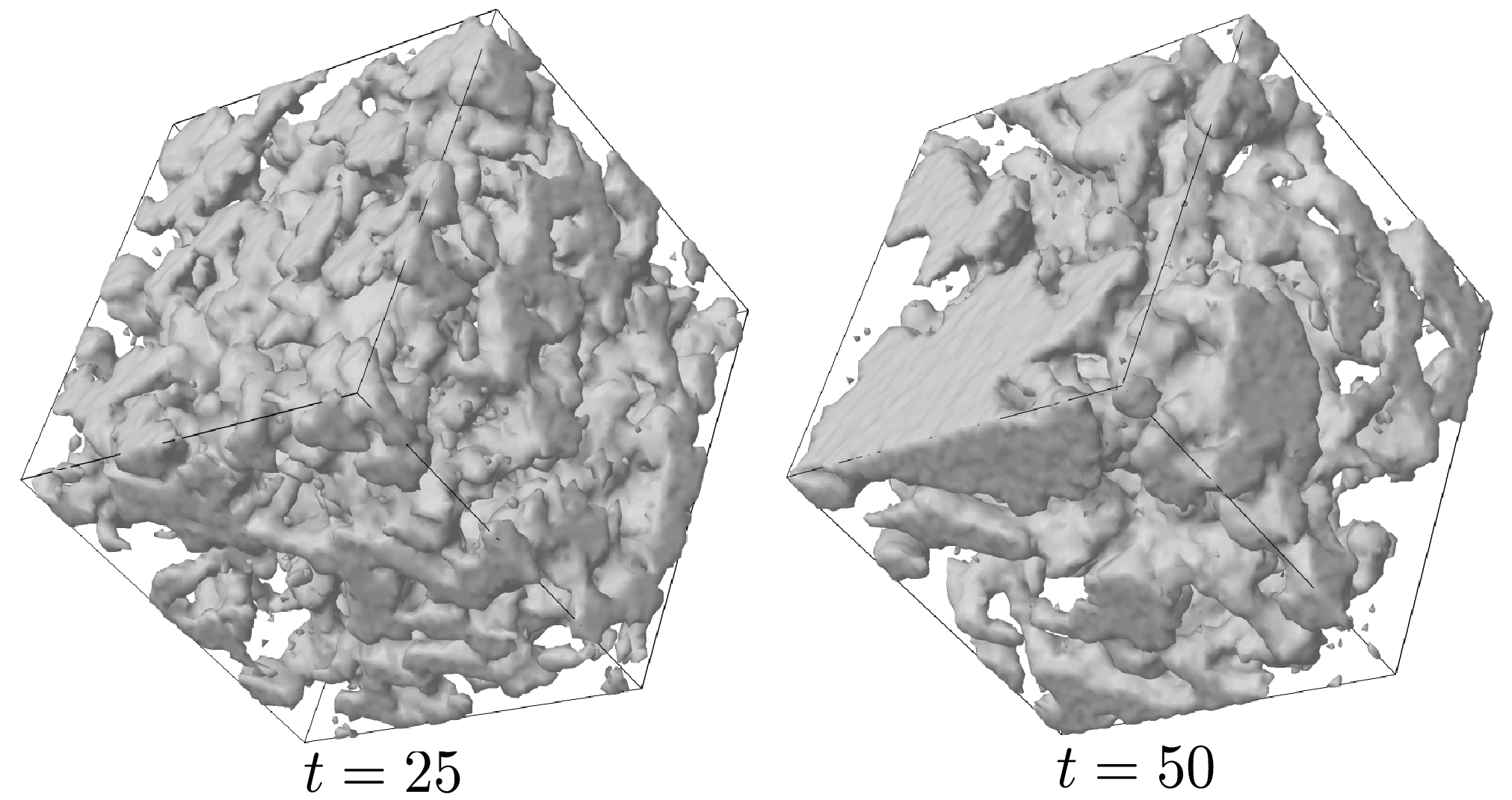}
    \caption{Surface plots of the 3D configurations at two different times corresponding to the evolution process in Fig. \ref{fig:layerview}. Snapshots are prepared using the `van der Waals surface' tool in JMOL \cite{jmol, jmolsurfacedoc}.}
    \label{fig:surface}
\end{figure}

First we present the results for the conserved momentum dynamics. 
In Fig. \ref{fig:layerview}, the two-dimensional (2D) cross-sectional views 
of snapshots at different times, starting from $t=0$, i.e., the instant of quench,
have been presented for a system of size $L=100$. 
As the system is quenched to $T=0.5$, that falls in the coexistence region of 
even the passive model, for the chosen density, 
particle-rich and particle-poor domains form and grow with time. 
Note that the used alignment activity is known to broaden the coexistence region. 
Clearly phase separation is observed.
Though these 2D cross-sections provide an impression that the domains are disconnected, 
from careful examination of the structure \cite{jmol, jmolsurfacedoc} in $d=3$, 
it appears that there exists interconnectedness. 
This can be appreciated from Fig. \ref{fig:surface}.
As in the case of 2D views, in Fig. \ref{fig:layerview}, 
the domain growth is clear from these snapshots as well. 

\begin{figure}[b]
    \centering
    \includegraphics[width=0.48\textwidth]{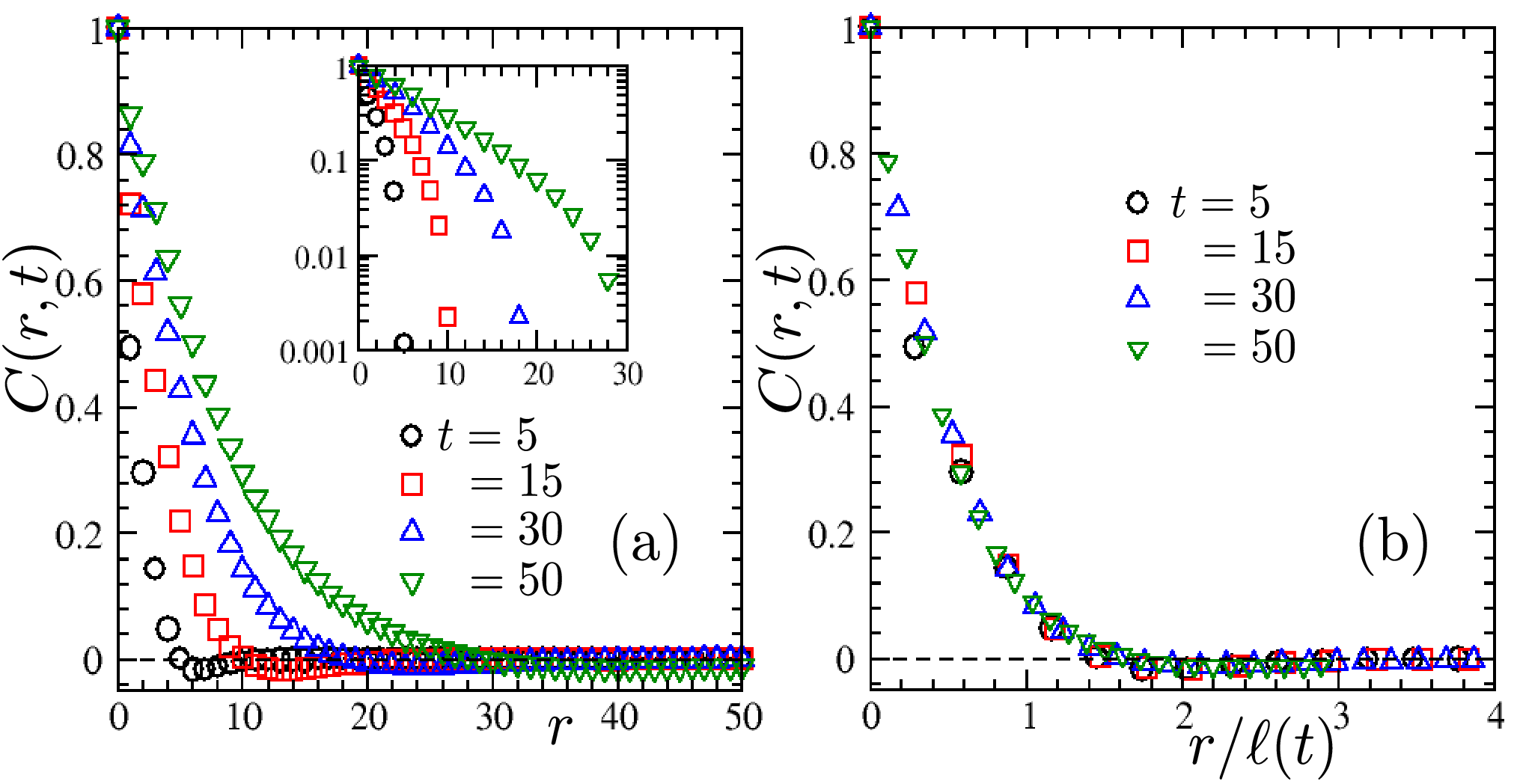}
    \caption{(a) The two-point equal-time correlation functions, for $L=100$, are plotted against $r$, at different times, for the conserved case. Data sets are normalized to set $C(0,t)=1$. Inset contains the same data sets on a semi-log scale. (b) Same as in the main frame of (a) but here the distance is scaled by $\ell$ at respective times. }
    \label{fig:cr}
\end{figure}

For a quantitative understanding of the structural evolution,
we examine the behavior of $C(r,t)$ in Fig. \ref{fig:cr}, from different times. 
For the calculation of $C(r,t)$, a continuum system at a given time, $t$, 
has been mapped onto a simple cubic lattice of lattice constant $\sigma$. 
For a lattice point, located at $\vec{r}$, 
a local density $\rho(\vec{r},t)$ has been calculated 
considering only nearest-neighbors and a local order-parameter
$\psi(\vec{r},t)$ has been assigned to it in a way that $\psi(\vec{r},t) = +1$ 
if $\rho(\vec{r},t) > \rho_0$, $-1$ otherwise. 
Nearly exponential decay of $C(r,t)$, 
till the decay of $C(r,t)$ to almost $90\%$ of its maximum value, 
is suggested by the plots, in a semi-log scale, in the inset of Fig. \ref{fig:cr}(a). 
Furthermore, a nice collapse of data is seen in Fig. \ref{fig:cr}(b), 
where the $r$-axis is scaled by the characteristic lengths $\ell(t)$. 
Here, at a given $t$, $\ell(t)$ is equal to the distance at which 
$C(r,t)$ falls off to a reference value $C_{\rm{ref}}=0.1$. 
This time-invariant nature of the structure, 
as implied by the presence of such scaling behavior, 
is consistent with a power-law growth of $\ell(t)$ with the increase of $t$.

\begin{figure}[h]
    \centering
    \includegraphics[width=0.48\textwidth]{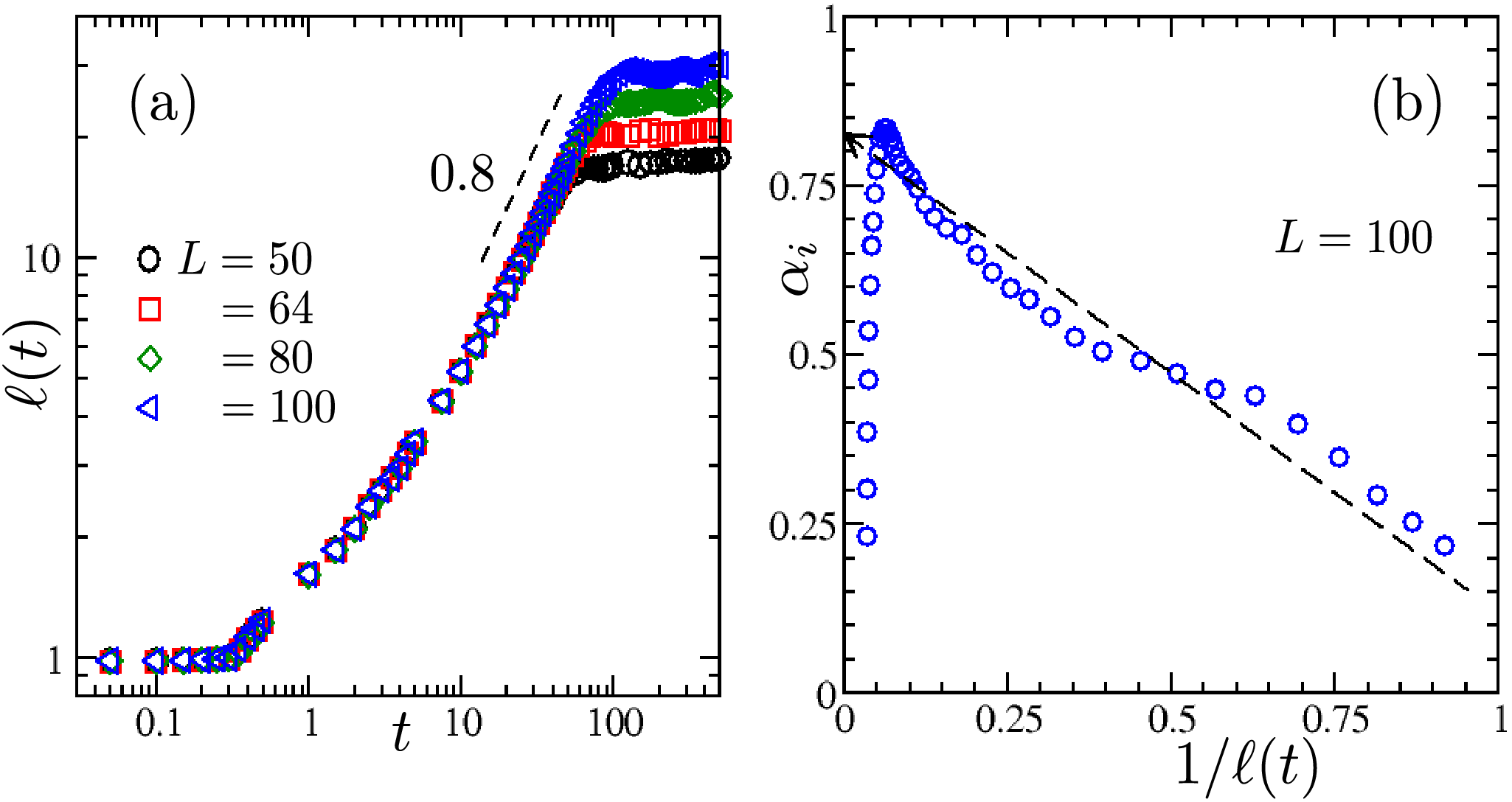}
    \caption{(a) Plots of $\ell(t)$, as a function of $t$, for the conserved case, in a double-log scale, for several different system sizes. The dashed line represents a power law with an exponent of $0.8$. (b) A plot of the instantaneous exponent, $\alpha_i$, for the $L=100$ data in (a), as a function of $1/\ell(t)$. We have performed a running averaging to obtain a smooth data set. The arrowheaded dashed line is a linear guide to the eye towards the limit $\ell \to \infty$.}
    \label{fig:lcf_alpha}
\end{figure}

To capture the growth behavior quantitatively, 
$\ell(t)$ has been plotted as a function of $t$, 
in Fig. \ref{fig:lcf_alpha}(a), on a log-log scale. 
Data for different system sizes show a very nice overlap initially. 
With the increase of time, 
results from smaller systems keep departing because of finite-size effects.
The dashed line in Fig. \ref{fig:lcf_alpha}(a) corresponds to a power law having an exponent $0.8$. 
As the data sets have continuous curvature and results suffer from finite-size effects, 
an accurate determination of the exponent requires more careful analysis. 
In presence of curvature, a better estimation of this exponent can be done 
from the calculation of the instantaneous exponent, $\alpha_i$, 
which is defined as \cite{huse1986corrections, amar88}
\begin{equation}
    \alpha_i = \dv{\ln \ell}{\ln t}. \label{eq:alpha_i}
\end{equation}
A plot of $\alpha_i$, as a function of $1/\ell$, is shown in Fig. \ref{fig:lcf_alpha}(b). 
The dashed line with an arrowhead here is an extrapolation of the data set 
in the limit $\ell \to \infty$, which indicates a value of $\alpha\simeq 0.825$. 
The downward bending of the data near small $1/\ell$, i.e., large $\ell(t)$, 
is related to the finite-size effects. 

The limitation due to finite size of systems we exploit to an advantage, 
to estimate $\alpha$ via the FSS analysis below. 
Finiteness respects the growth of domains up to only a finite value, 
say $\ell_{\rm{max}}$. 
After the quench, during the growth process, 
a system takes a certain time, $t_0$, to enter the self-similar growth regime. 
Taking this latter fact into account, Eq. (\ref{eq:growth}) can be modified as 
\cite{majumder11}
\begin{equation}
    \ell(t) = \ell_0 + A(t-t_0)^\alpha, \label{eq:modgrowth}
\end{equation}
where $\ell_0= \ell(t_0)$. 
The late time effect can be incorporated by writing 
\begin{equation}
    \ell(t)-\ell_0=Y(y)(\ell_{\rm{max}}-\ell_0), \label{eq:FSS}
\end{equation}
where $Y(y)$ is a scaling function independent of system size, 
depending only on the dimensionless scaling variable \cite{majumder11}
\begin{equation}
    y = {\frac{(\ell_{\rm{max}}-\ell_0)} {(t-t_0)}} ^ {1/\alpha}. \label{eq:y}
\end{equation}
This implies, if $Y(y)$ is plotted as a function of $y$, 
with correct choices of $\ell_0$, $t_0$, and $\alpha$, 
data from different system sizes should exhibit a collapse. 
Furthermore, in the limit $y\to \infty$, i.e., at early time, 
it is expected to exhibit the power-law behavior $Y(y) \sim y^{-\alpha}$ 
so that Eq. \eqref{eq:modgrowth} is obeyed. 

\begin{figure}[h]
    \centering
    \includegraphics[width=0.48\textwidth]{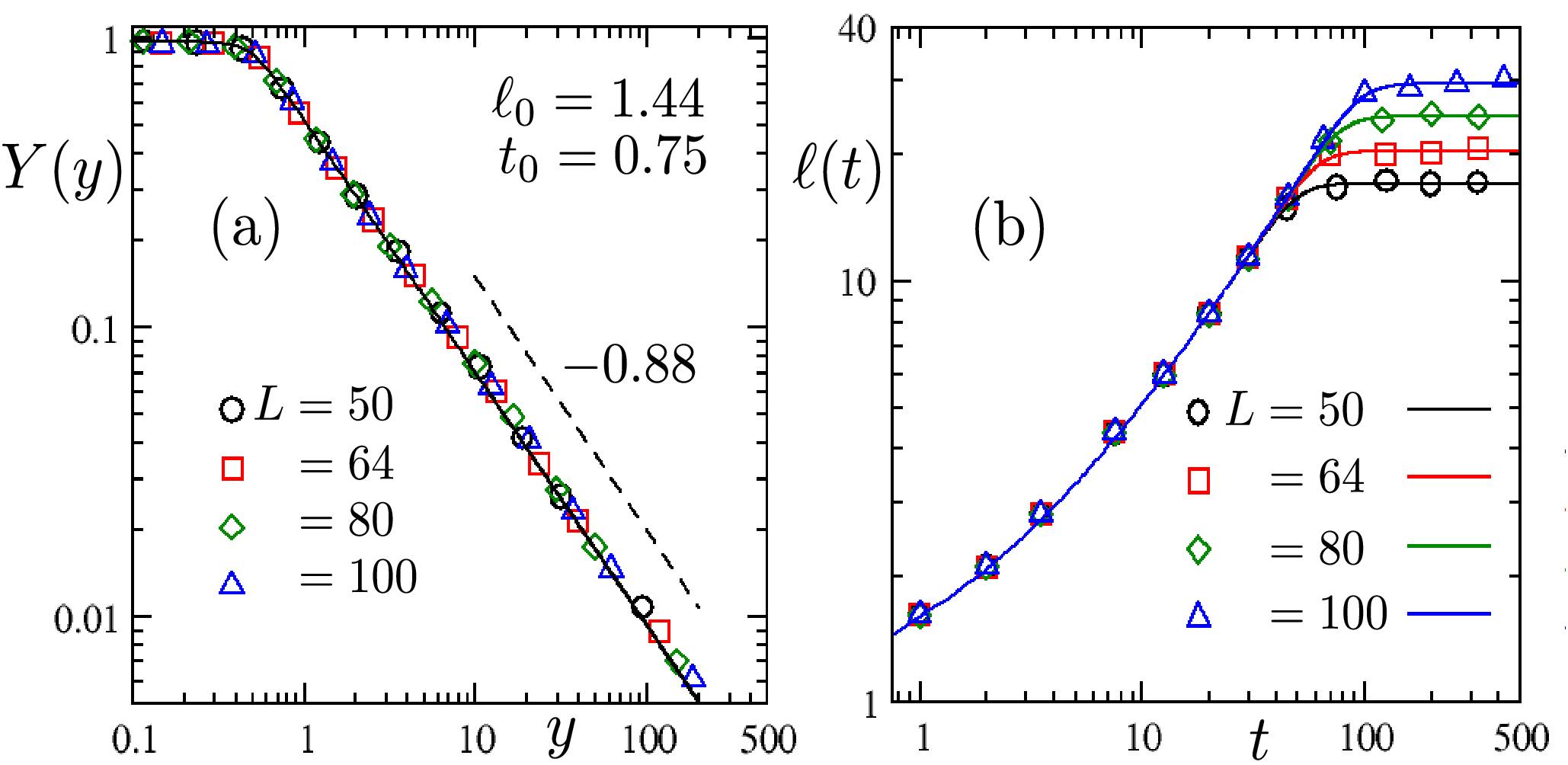}
    \caption{(a) Plot of the scaling function, $Y(y)$, as a function of $y$, in a double-log scale, for the conserved case. To obtain the collapse, we have used data from several system sizes, as mentioned in the figure. This was achieved for $t_0=0.75$, $\ell_0=1.44$, and $\alpha=0.88$. The dashed line represents a power law decay of $Y(y)$, for large $y$, with an exponent $0.88$. The solid line is a fit of the combined data set to the function in Eq. \eqref{eq:Yy_fn}. (b) Plots of Eq. \eqref{eq:growthfn}, along with the growth data sets, $t_0$ onwards, for different $L$. We have thinned out the data for the visibility of analytical lines.}
    \label{fig:Yofy}
\end{figure}

In Fig. \ref{fig:Yofy}(a), $Y(y)$ has been plotted as a function of $y$, 
on a log-log scale, for the choices $\ell_0=1.44$, $t_0=0.75$ and $\alpha=0.88$, 
for all the considered system sizes. 
There is a nice collapse of the data sets. 
As shown in the figure by a dashed line, 
$Y(y)$ obeys a power-law with an exponent $\alpha=0.88$ for large values of $y$. 
There is a crossover to another form as $y$ decreases before it reaches a constant. 
A function to describe data on either side of the crossover was obtained recently 
in a study of disease spread \cite{das21}. 
In fact, the construction was motivated by the finite-size scaling behavior 
in the context of phase transition 
and applicability to the latter was demonstrated more recently in a simple case \cite{das2023finitesize}. 
This has the form
\begin{equation}
    Y(y) = Y_0\left(b+\dfrac{y^{\theta}}{\alpha}\right)^{-\alpha/\theta}, \label{eq:Yy_fn}
\end{equation}
where $Y_0$, $b$ and $\theta$ are positive constants. 
From Eq. \eqref{eq:Yy_fn}, it follows that for $y \to \infty$, $Y(y) \sim y^{-\alpha}$, 
as expected from the construction of the scaling. 
With the above mentioned choice of $\alpha$, viz., $0.88$, we have obtained 
a very good fit to the collapsed data sets, presented in Fig. \ref{fig:Yofy}(a). 
The function has been shown there with a solid line. 
The best fit values of the other three parameters are 
$Y_0 \simeq 0.53$, $b \simeq 0.01$, and $\theta \simeq 6.50$. 
Here note that $\theta$ determines the sharpness of the crossover and 
can be treated as a parameter determining universality classes for finite-size effects.
Combining Eqs. \eqref{eq:FSS}, \eqref{eq:y} and \eqref{eq:Yy_fn}, one can write 
\begin{equation}
    \ell(t) = \ell_0 + Y_0 (\ell_{\rm{max}}-\ell_0) \left[b + \dfrac{(\ell_{\rm{max}}-\ell_0) ^ {\theta/\alpha}}{\alpha (t-t_0)^\theta} \right] ^ {-\alpha / \theta}. \label{eq:growthfn}
\end{equation}
By calculating $\ell_{\rm{max}}$ for different $L$, 
as the mean values at the saturations of $\ell(t)$, 
the function in Eq. \eqref{eq:growthfn} has been plotted for several $L$ 
with continuous lines in Fig. \ref{fig:Yofy}(b). 
For each $L$, this function nicely passes through the corresponding data set, 
i.e., Eq. \eqref{eq:growthfn} is appropriately describing the growth behavior 
of all the finite systems. 

\begin{figure}[h]
    \centering
    \includegraphics[width=0.3\textwidth]{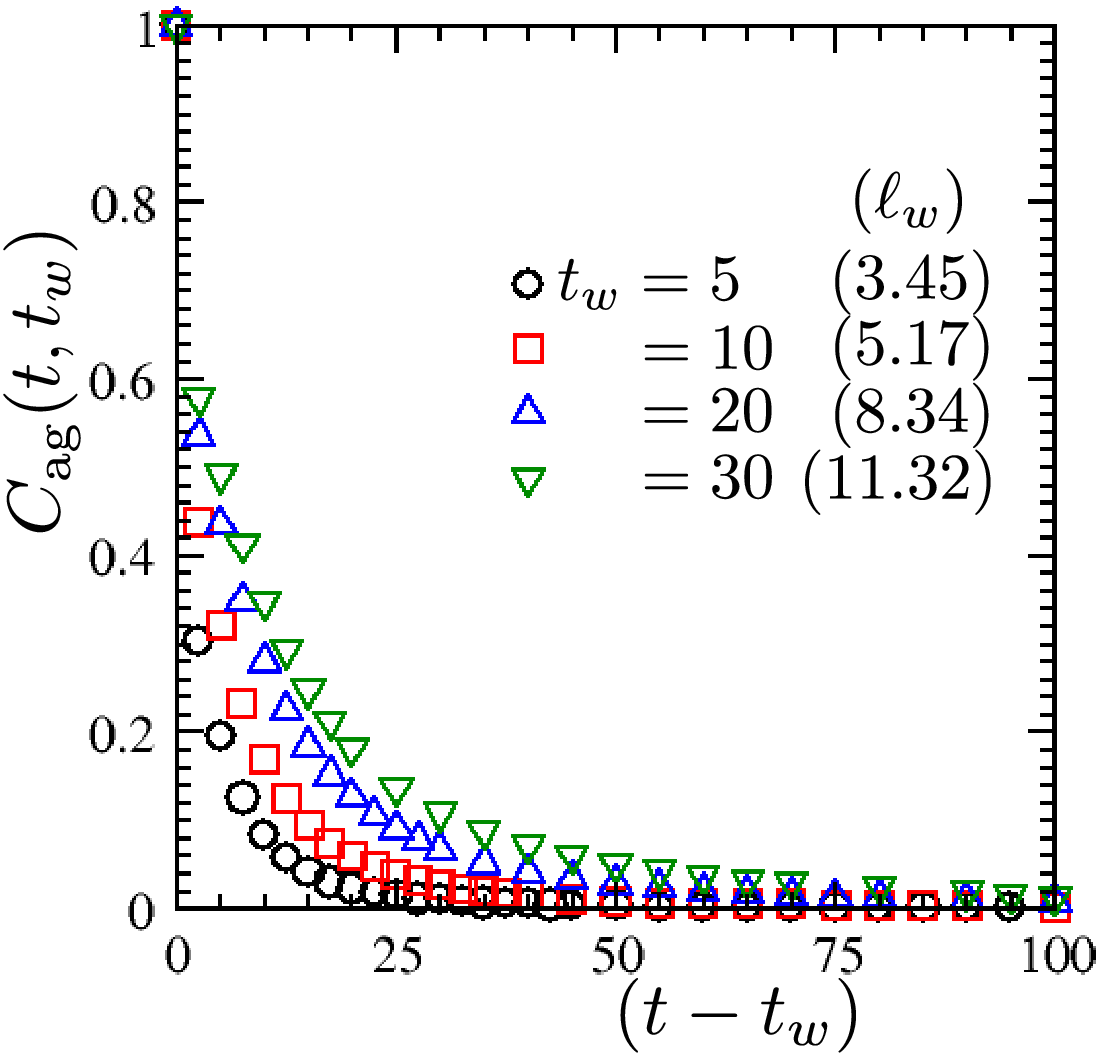}
    \caption{Plots of $C_{\rm{ag}}(t,t_w)$, the autocorrelation function, vs. $(t-t_w)$, for different ages ($t_w$) of the system with $L=100$ and conserved velocity. Corresponding $\ell_w$ values are provided inside the parentheses.}
    \label{fig:cag}
\end{figure}

\begin{figure}[h]
    \centering
    \includegraphics[width=0.48\textwidth]{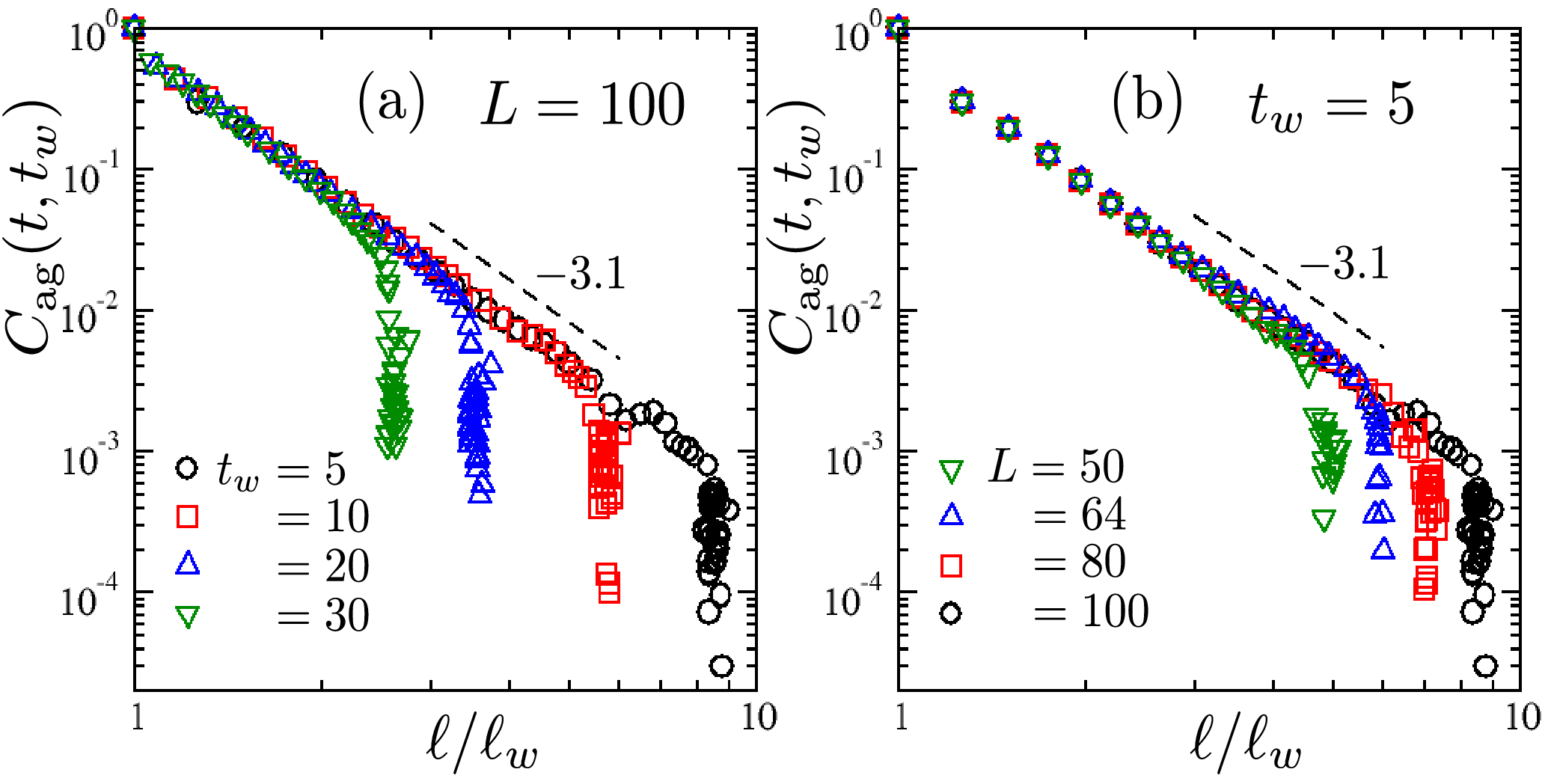}
    \caption{Double-log plots of $C_{\rm{ag}}(t,t_w)$ as a function of $\ell/\ell_w$ for the conserved case: (a) for fixed system size and different $t_w$, as in Fig. \ref{fig:cag}; and (b) for different system sizes for a single $t_w$. The dashed lines represent power-law falls.}
    \label{fig:cag_lbylw_comp}
\end{figure}

Next, we focus on the aging properties. 
In Fig. \ref{fig:cag}, we plot $C_{\rm{ag}}(t,t_w)$ 
with the variation of the shifted time $(t-t_w)$, for different $t_w$ values. 
Here, to capture the correlation between two systems conveniently, 
one at $t_w$ and another at $t$, 
the configuration at an instant has been simply mapped onto a simple cubic lattice 
of lattice constant $\sigma$, without any coarse-graining, unlike for the calculation of $C$. 
For a given $t_w$, the correlation between the two systems decreases with increasing $t$, as expected. 
Furthermore, as the system ages, i.e., $t_w$ increases, the relaxation becomes slower. 
Such an absence of time-translation invariance is also expected for evolving systems. 
But there typically exists a scaling, as described previously, 
which can be seen if $C_{\rm{ag}}(t,t_w)$ is plotted vs $\ell/\ell_w$. 
This is demonstrated in Fig. \ref{fig:cag_lbylw_comp}(a). 
A good quality collapse of data sets for different $t_w$ can be appreciated; 
deviation in the tail part of each data set has its origin in the finite size of a system. 
In a log-log scale, the finite-size unaffected data appear to have a power-law decay, 
as noted in Eq. \eqref{eq:agingexp}. 
The dashed line in Fig. \ref{fig:cag_lbylw_comp}(a) indicates that 
the decay exponent $\lambda$ has a value $\simeq 3.1$. 
Fig. \ref{fig:cag_lbylw_comp}(b) is similar to Fig. \ref{fig:cag_lbylw_comp}(a), 
but here $C_{\rm{ag}}(t,t_w=5)$ has been plotted as a function of $\ell/\ell_w$, 
for different $L$. 
Here too, data sets for all the system sizes appear to collapse onto a master curve, 
$\lambda$ remaining unchanged, of course. 
Similar (finite-size) features in the two parts, (a) and (b), have the following reason. 
When the system size is kept fixed, for increasing $t_w$, 
a system has lesser further opportunity to grow. 
Next, we extend our investigation to the study of finite-size scaling in aging, 
presence of which is demonstrated in a few passive situations \cite{midya14, midya15, vadakkayil19}.

\begin{figure}[h]
    \centering
    \includegraphics[width=0.48\textwidth]{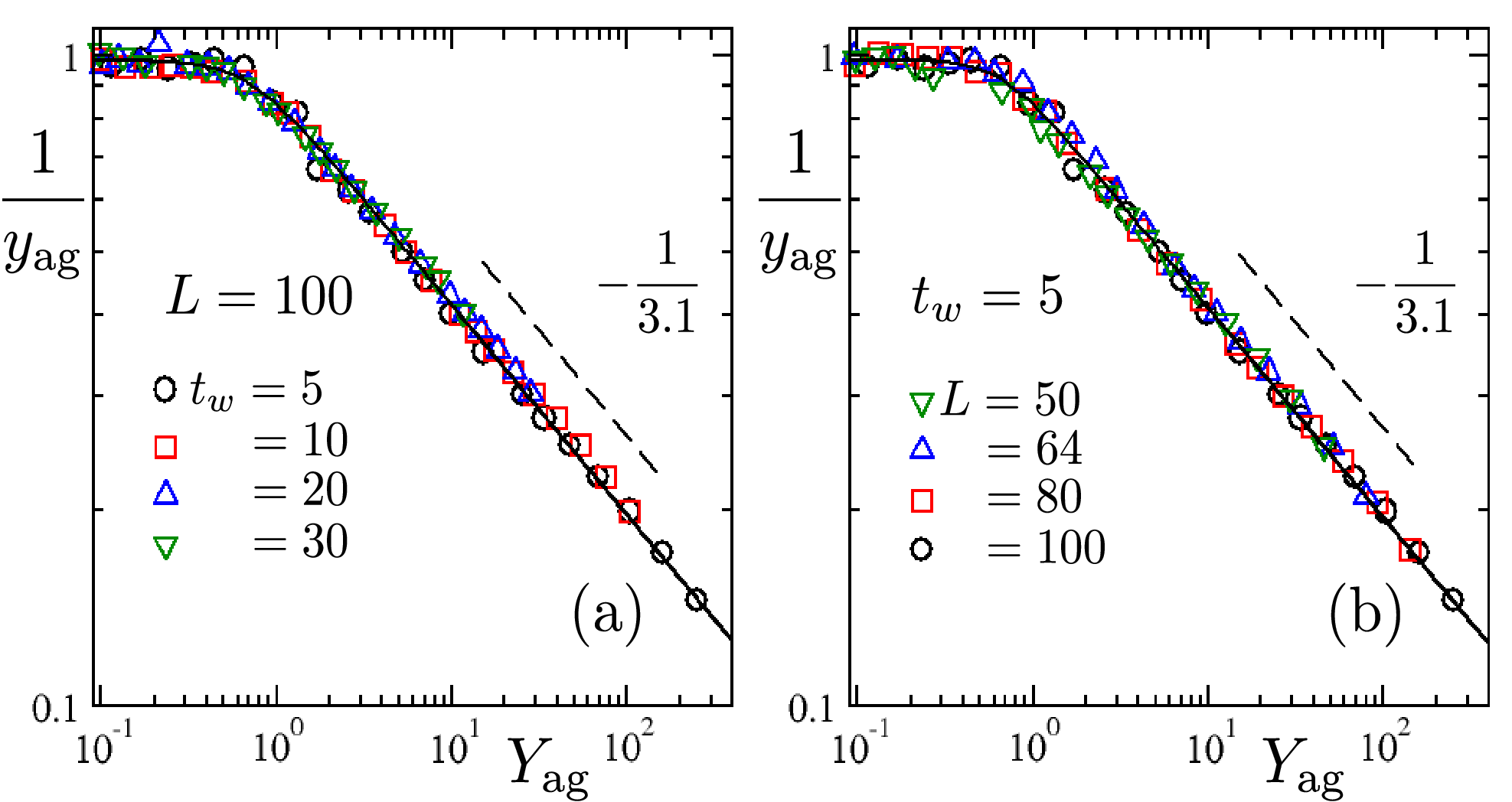}
    \caption{Plots of $1/y_{\rm{ag}}$ vs. $Y_{\rm{ag}}$ in double-log scales, using data sets of Fig. \ref{fig:cag_lbylw_comp}: (a) for a fixed system size, with different $t_w$ values, and (b) for different system sizes, with $t_w=5$. The dashed lines are power-laws with an exponent $-1/3.1$. The solid lines are fits of the combined data sets to the function in Eq. \eqref{eq:Yag_fn}.
    }
    \label{fig:Yag}
\end{figure}

A dimensionless scaling variable, in this case, 
can be taken as \cite{midya14, midya15, vadakkayil19} 
$y_{\rm{ag}}=\ell_{\rm{max}}/\ell$. 
Then, in the limit $y_{\rm{ag}} \to 0$, i.e., $\ell \to \infty$, 
data from systems with finite $L$ will deviate from their thermodynamic limit behavior. 
A scaling function, $Y_{\rm{ag}} (y_{\rm{ag}})$, 
that remains invariant for any system size, can be written as
\begin{equation}
    Y_{\rm{ag}}(y_{\rm{ag}}) = C_{\rm{ag}}(t,t_w) \left( \frac{\ell_{\rm{max}}}{\ell_w} \right) ^ \lambda. \label{eq:Ycag}
\end{equation}
Note that here we have discarded a correction term that was included in Ref. \cite{midya14, midya15, vadakkayil19}. 
This is because of the linear appearance of the plots on a log-log scale from the very beginning.
Such a construction should allow one to obtain overlap of data 
from different $L$ for a fixed value of $t_w$, 
as well as for different $t_w$, for a fixed $L$.
As $C_{\rm{ag}}$ follows the scaling property noted in Eq. (\ref{eq:agingexp}),
when $\ell \gg \ell_w$, in the limit $L(\mbox{or}~\ell_{\rm{max}}) \to \infty$, i.e., $y_{\rm{ag}} \to \infty$, 
$Y_{\rm{ag}}$ should follow the power law $Y_{\rm{ag}} \sim y_{\rm{ag}}^\lambda$, i.e.,
\begin{equation}
    \dfrac{1}{y_{\rm{ag}}} \sim Y_{\rm{ag}}^{-1/\lambda}. \label{eq:Yaglimiting}
\end{equation}
Using the data sets in Fig. \ref{fig:cag_lbylw_comp}(a), 
$1/y_{\rm{ag}}$ has been plotted as a function of $Y_{\rm{ag}}$, 
for different choices of $t_w$, in Fig. \ref{fig:Yag}(a), in a log-log scale. 
A nice data collapse is observed for the choice of $\lambda=3.1$. 
In Fig. \ref{fig:Yag}(b), the same has been done 
with the data sets in Fig. \ref{fig:cag_lbylw_comp}(b). 
A good data collapse has been obtained again for $\lambda=3.1$. 
As shown by the dashed lines, Eq. \eqref{eq:Yaglimiting} is correctly describing 
the scaling in the limit $y_{\rm{ag}} \to \infty$ with an exponent $=-1/3.1$. 
For other values of $t_w$ also, the same exercises have been carried out 
(the results are not presented here). 
In all cases, the same value of $\lambda$ is seen to be a good choice.

The collapsed data in Fig. \ref{fig:Yag} 
appear to have a very similar look as those for the case of FSS in length scale 
(see Fig. \ref{fig:Yofy}(a)). 
Thus, one should look for a fit to the form 
\begin{equation}
    \dfrac{1}{y_{\rm{ag}}} = A_0 \left( B + \dfrac{Y_{\rm{ag}}^{\phi}}{q} \right) ^ { - q / \phi}. \label{eq:Yag_fn}
\end{equation}
The best fits, in both cases, are obtained for the choice $q=1/3.1$, 
which we take as the value of $1/\lambda$. 
Other parameters are $A_0 \simeq 0.99$, $B \simeq 1.04$ 
and $\phi \simeq 2.79$ for the case of Fig. \ref{fig:Yag}(a), 
and $A_0 \simeq 0.96$, $B \simeq 0.81$ and $\phi \simeq 3.21$ 
for the case of Fig. \ref{fig:Yag}(b). 

\begin{figure}[h]
    \centering
    \includegraphics[width=0.48\textwidth]{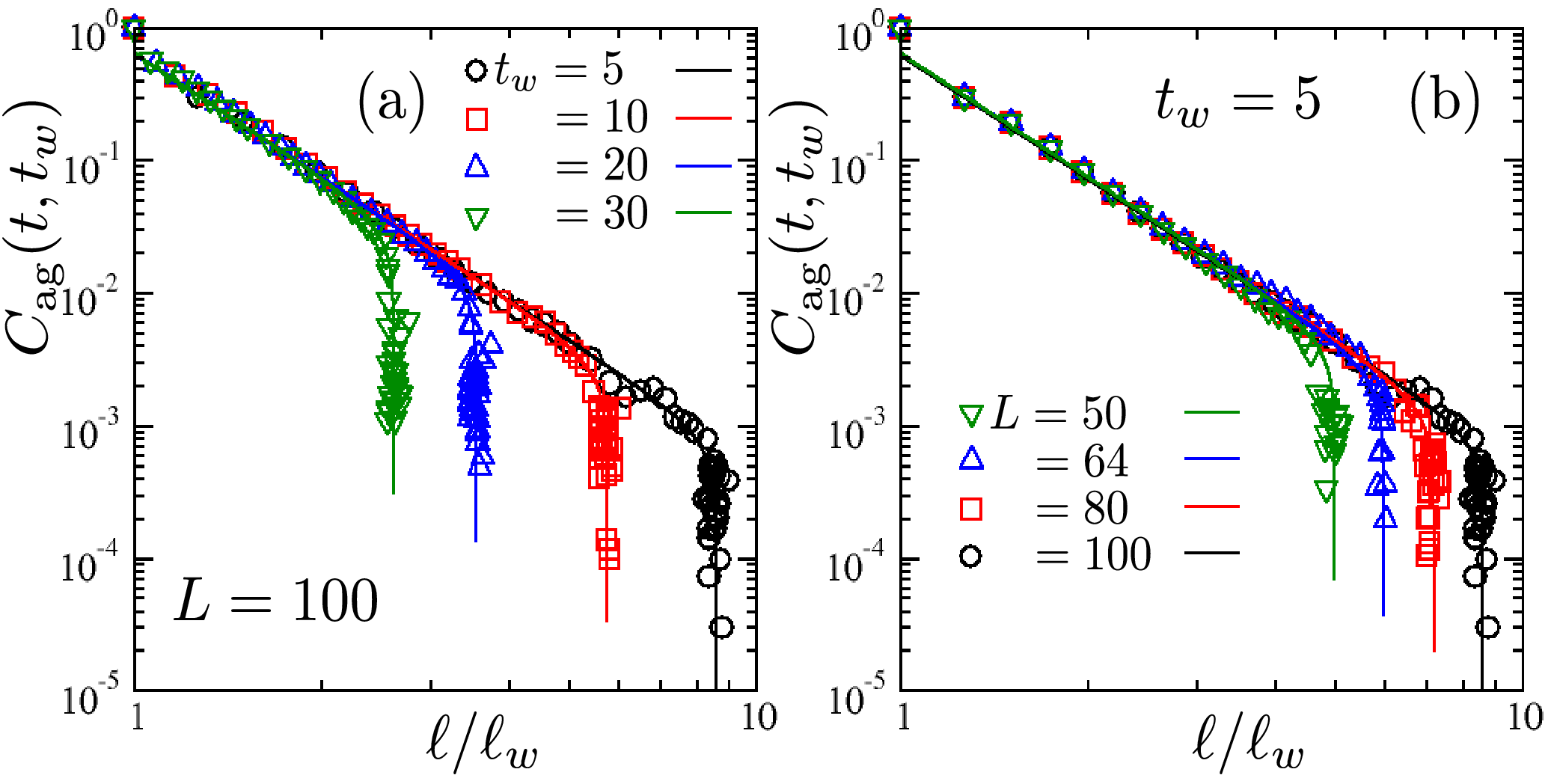}
    \caption{Plots of Eq. \eqref{eq:cagfnform} (see the continuous lines) along with the $C_{\rm{ag}}(t,t_w)$ data for (a) different $t_w$, with $L=100$, and (b) different $L$, with $t_w=5$. These results are for conserved velocity.}
    \label{fig:cag_fnfit}
\end{figure}

Combining Eqs. \eqref{eq:Ycag} and \eqref{eq:Yag_fn}, 
like in the case of domain growth, 
one can write a system-size dependent functional form of $C_{\rm{ag}}(t,t_w)$ as
\begin{equation}
    C_{\rm{ag}}(t,t_w) = \left[ \left\{ \left( \dfrac{x}{A_0 X} \right) ^{-\lambda\phi} - B \right\} \dfrac{1}{\lambda} \left( \dfrac{\ell_w}{\ell_{\rm{max}}} \right) ^{\lambda\phi} \right] ^{1/\phi}, \label{eq:cagfnform}
\end{equation}
where $x= \ell / \ell_w$ and $X= \ell_{\rm{max}} / \ell_w$. 
In Eq. \eqref{eq:cagfnform}, we have replaced $ y_{\rm{ag}} $ 
by $ \ell_{\rm{max}}/\ell $.  
Incorporating the values of the parameters $A_0$, $B$, and $\phi$, 
as obtained from the fitting exercise, 
the function in Eq. \eqref{eq:cagfnform} has been compared with the simulation data 
for each of the cases presented in Fig. \ref{fig:cag_lbylw_comp}. 
See Fig. \ref{fig:cag_fnfit}. 
The function appears to describe the simulation data sets quite well, once again.

\begin{figure}[hbt!]
    \centering
    \includegraphics[width=0.48\textwidth]{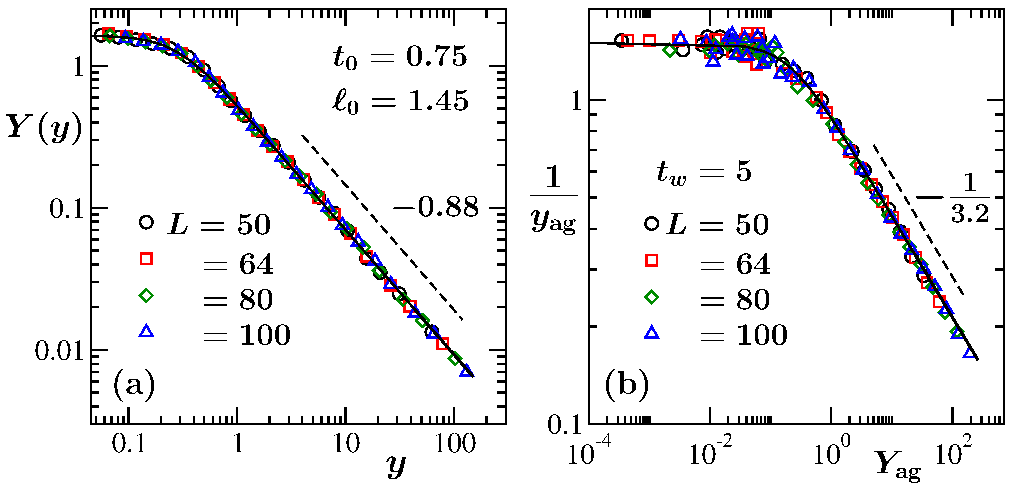}
    \caption{(a) A double-log plot of $Y(y)$, vs $y$, for domain length, $\ell(t)$, showing overlap of data for different $L$, with non-conserved momentum. This can be compared with Fig. \ref{fig:Yofy}(a). The continuous line here is a fit of the collapsed data to Eq. \eqref{eq:Yy_fn} and the dashed line is a power-law corresponding to $\alpha=0.88$. 
    (b) A plot of $1/y_{\rm{ag}}$ vs $Y_{\rm{ag}}$, in a double-log scale, for different $L$, with $t_w = 5$, like in Fig. \ref{fig:Yag}(b). The dashed line is a power law representing $\lambda=3.2$ and the solid line is a fit of the collapsed data to Eq. \eqref{eq:Yag_fn}. 
    For both (a) and (b), in the scaling functions in Eqs. \eqref{eq:Yy_fn} and \eqref{eq:Yag_fn}, instead of $\ell_{\rm{max}}$, we have used $\ell_{\rm{dev}}$, which is the length at which the deviation from the power-law behavior of $\ell(t)$ starts. The behavior of $\ell_{\rm{dev}}$ is also linear with the system size, as in the case of $\ell_{\rm{max}}$.}
    \label{fig:fss_ncop}
\end{figure}

\begin{figure}[hbt!]
    \centering
    \includegraphics[width=0.48\textwidth]{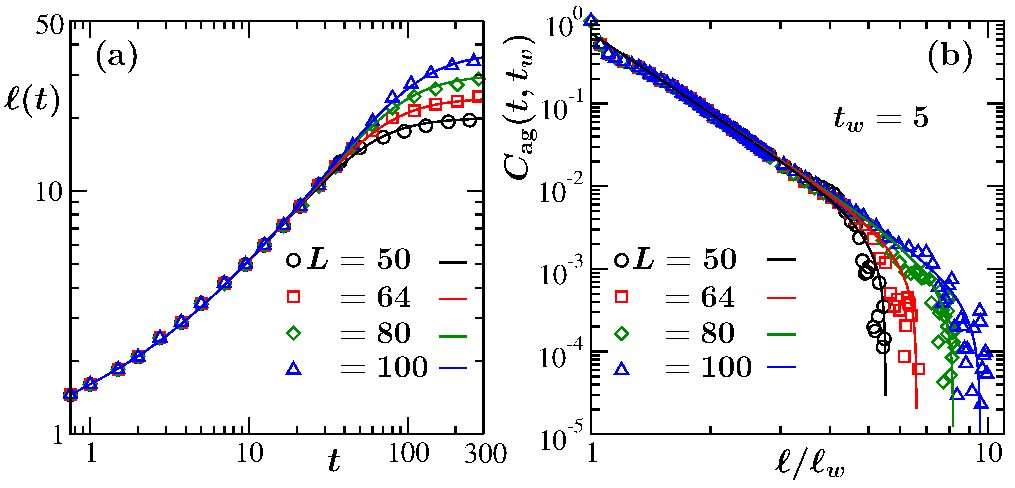}
    \caption{(a) Plots of $\ell(t)$ as a function of $t$, in a double-log scale, for different $L$, in the case of non-conserved momentum. The continuous lines here correspond to Eq. \eqref{eq:growthfn}. 
    (b) Plots of $C_{\rm{ag}}(t,t_w)$ vs $\ell/\ell_w$, from different $L$, for $t_w = 5$, along with the functional form (solid lines) given in Eq. \eqref{eq:cagfnform}. 
    These plots can be compared with Figs. \ref{fig:Yofy}(b) and \ref{fig:cag_fnfit}(b).}
    \label{fig:fitting_ncop}
\end{figure}

Finally, we compare some of the above results with the case of non-conserved momentum 
dynamics. 
Scaling of $C(r,t)$ and $C_{\rm{ag}}(t,t_w)$ have similar behavior 
as in the case of conserved momentum dynamics. 
Thus, we present directly the nontrivial finite-size scaling plots 
via which we extract the exponents $\alpha$ and $\lambda$. 
Fig. \ref{fig:fss_ncop}(a) is similar to the Fig. \ref{fig:Yofy}(a), 
where we show the scaling plots, using data from different system sizes, 
for the domain length, in a log-log scale. 
The functional form in Eq. \eqref{eq:Yy_fn} fits well to the data sets 
and the value of the exponent $\alpha \simeq 0.88$ 
matches nicely with the case of conserved momentum dynamics. 
The numbers for the other fitting parameters are $Y_0 \simeq 0.56$, $b \simeq 0.06$, 
and $\theta \simeq 2.28$. 
Furthermore, in Fig. \ref{fig:fss_ncop}(b) we present the results for 
finite-size scaling of $C_{\rm{ag}}(t,t_w)$, 
using data from different system sizes for $t_w = 5$. 
This can be compared with Fig. \ref{fig:Yag} (b). 
Here also the data sets are in good agreement with the functional form in 
Eq. \eqref{eq:Yag_fn}, for the exponent $\lambda \simeq 3.2$, 
which is within about $3\%$ from the value obtained for the conserved case. 
The values of the other fitting parameters are the following: 
$A_0 \simeq 1.18$, $B \simeq 0.37$, and $\phi \simeq 1.33$. 
From these analyses, we conclude that the dynamics is only weakly dependent 
on whether the momentum is conserved or not. Finally, in Figs. \ref{fig:fitting_ncop} (a) in (b) we show the comparisons of simulation data with the analytical forms for different systems sizes. Agreements appear to be very satisfactory again.

While the thermodynamic limit exponents appear almost same 
for the conserved and the nonconserved cases, 
from the appearances of the plots as well as 
the best fit values of the parameters $\theta$ and $\phi$, 
it is evident that the crossover from the thermodynamic limit 
to the finite-size behavior is sharper in the conserved case. 
From the plots of domain growth, in the finite-size limit oscillatory behavior is clear. 
This is related to density fluctuations within domains with time 
in the steady-state situations, a sort of domain breathing. 
This gets reflected in the aging case also. 
When the simulation data are plotted with lines, this feature becomes clear.  
Even though the above-mentioned fluctuation in the domain size appears to grow larger 
with the increase of system size, it does not truly diverge. 
Thus, when calculated after averaging over simulation runs 
with adequately large number of initial configurations, 
it is expected that the oscillations will disappear for all $L$.

\section{Conclusion}\label{sec:conclusion}

We study the kinetics of evolution in active matter systems 
using a model where the activity rule is Vicsek-like \cite{vicsek95}. 
The (overlap preventing) inter-particle passive interaction is introduced via a 
variant \cite{roy13JCP, midya17} 
of the standard Lennard-Jones potential \cite{allenbook}. 
The density of particles within the systems is taken to be close to 
the (vapor-liquid) critical value corresponding to the pure passive limit of the model. 

The dynamics of the active systems, as they approach the steady states, 
has similarities with passive systems \cite{katyal20,gonnella15} 
at the qualitative level, including the finite-size scaling (FSS) behavior. However,
the addition of the active alignment interaction 
brings quantitative changes in all aspects of evolution. 
During the relaxation of the systems, we observe scaling of 
the two-point equal-time correlation function, $C(r,t)$, 
implying the satisfaction of self-similarity of the structures at different times. 
We calculated the domain length, $\ell(t)$, from the above scaling property 
and estimated the growth exponent, $\alpha$, 
from the FSS analysis of $\ell(t)$ considering different system sizes. 
We also obtain a system-size dependent functional form of $\ell(t)$, 
using a recently constructed FSS function. 
The value of $\alpha$ has been found to be $\simeq 0.88$, 
which is higher than that in the passive case without hydrodynamics 
\cite{lifshitz1961kinetics,das17,siggia1979}. 
In the active case, the presence of alignment interaction 
could give rise to an advection-like effect, accelerating cluster growth. 
The aging exponent, $\lambda$, calculated from the autocorrelation function, 
$C_{\rm{ag}}(t,t_w)$, has a value $\simeq 3$, 
smaller than a value for nonhydrodynamic passive case \cite{midya15}. 
This is consistent with the fact of stronger aging for faster growth. 
In this case also, we observed very good FSS behavior, 
analysis via which allowed us the estimation of the corresponding exponent accurately. 
The system-size dependent functional form of 
$C_{\rm{ag}}(t,t_w)$ is obtained here as well. 
We also find that the values of the exponents $\alpha$ and $\lambda$ 
do not depend strongly on whether the dynamics preserves the total momentum. 
However, the effects of conservation are present in the finite-size effects.

It is important to make a few remarks on the finite-size behavior here. 
From Figs. \ref{fig:Yofy}(a) and \ref{fig:fss_ncop}(a) it is clear that at large $y$, 
$Y$ exhibits a power-law behavior, with exponent $\alpha$. 
As $y\to 0$, $Y$, on the other hand, tends to a constant value. 
The departure from one limit to the other may happen in different fashions. 
The construction of Eq. \eqref{eq:Yy_fn} was done \cite{das21} by considering 
a power-law form of the deviation, characterized by an exponent $\theta$. 
High quality fit of the simulation data sets with the thus obtained form of $Y$ 
verifies the accuracy of the consideration. 
The same scenario applies for the aging behavior. 
It may not, however, be expected that for all models and physical situations 
a power-law form for the departure will apply. 
It will be interesting to obtain $Y$ 
by considering exponential and logarithmic forms for the above mentioned deviation. 
Then, in a broader sense, universalities in finite-size behavior can be classified 
based on the applicability of one or the other such function. 
Such universal features can be readily verified by studying other simple 
but nontrivial models and situations of current interest 
\cite{Chatterjee2020, dittrich23, patridge19}.

\section*{Acknowledgements:}\label{sec:acknowledgements}
SKD is thankful to the Department of Biotechnology, India, 
for partial financial support via Grant No. LSRETJNC/SKD/4539.
SKD and NV acknowledge partial support from the 
Science and Engineering Research Board, India, via Grant No. MTR/2019/001585. 
The authors are thankful to the supercomputing facility, 
PARAM Yukti, at JNCASR, under National Supercomputing Mission.


\bibliography{refActive}

\end{document}